\documentclass[runningheads]{llncs}
\usepackage{stfloats}
\fnbelowfloat
\usepackage[T1]{fontenc}

\usepackage{graphicx}
\usepackage{xcolor}
\usepackage{amsmath}
\begin{document}
\title{
  {\small \textcolor{red}{\textbf{Accepted to Appear in the Proceedings of}}\vspace{-1.5ex}\\
  \textcolor{red}{\textbf{16th International Conference on Learning Analytics \& Knowledge 2026}}}\\[1ex]
  Scaffolding Reshapes Dialogic Engagement in Collaborative Problem Solving:\\ Comparative Analysis of Two Approaches
}

\titlerunning{Scaffolding Reshapes Dialogic Engagement in Collaborative Problem Solving}
%
\author{Kester Wong\inst{1}\and
Feng Shihui\inst{2}\and
Sahan Bulathwela\inst{3}\and
Mutlu Cukurova\inst{1}}
\authorrunning{K. Wong et al.}
%
\institute{
$^{1}$ UCL Knowledge Lab, Institute of Education, University College London, UK\\
$^{2}$ The University of Hong Kong, Hong Kong, China\\
$^{3}$ AI Centre, Dept. of Computer Science, University College London, UK\\
\email{yew.wong.21@ucl.ac.uk}}
\maketitle              
\begin{abstract}
Supporting learners during Collaborative Problem Solving (CPS) is a necessity. Existing studies have compared scaffolds with maximal and minimal instructional support by studying their effects on learning and behaviour. However, our understanding of how such scaffolds could differently shape the distribution of individual dialogic engagement and behaviours across different CPS phases remains limited. This study applied Heterogeneous Interaction Network Analysis (HINA) and Sequential Pattern Mining (SPM) to uncover the structural effects of scaffolding on different phases of the CPS process among K-12 students in authentic educational settings. Students with a maximal scaffold demonstrated higher dialogic engagement across more phases than those with a minimal scaffold. However, they were extensively demonstrating scripting behaviours across the phases, evidencing the presence of overscripting. Although students with the minimal scaffold demonstrated more problem solving behaviours and fewer scripting behaviours across the phases, they repeated particular behaviours in multiple phases and progressed more to socialising behaviours. In both scaffold conditions, problem solving behaviours rarely progressed to other problem solving behaviours. The paper discusses the implications of these findings for scaffold design and teaching practice of CPS, and highlights the distinct yet complementary value of HINA and SPM approaches to investigate students' learning processes during CPS.
\end{abstract}

\section{Introduction}
Collaborative Problem Solving (CPS) involves two or more individuals communicating, collaborating, and incorporating components of cognition found in individual problem solving to reach a desired goal state \cite{oneil_2010ComputerBaseda,oecd_2017PISAb}. However, learners tend not to engage in such collaboration processes on their own \cite{weinberger_2007Scripting}. For young learners, this could be attributed to their overall lack of proficiency in CPS \cite{oecd_2017PISA}. Scaffolding has been explored extensively in education to address learners' zone of proximal development by organising learning to induce various developmental processes  (e.g., social, cognitive) \cite{vygotsky_1978Mind}. Hence, the use of scaffolding through various forms and principles of scripting has been widely explored in the Computer-Supported Collaborative Learning (CSCL) community to address the dynamic interaction between individuals and the collective group during CPS (e.g., \cite{kobbe_2007Specifying,fischer_2013Script}).

While existing studies and meta-reviews have highlighted the benefits of socio-cognitive scaffolding for increasing students' domain-specific knowledge and collaboration skills (e.g., \cite{vogel_2017SocioCognitive,miller_2024Comparing}), analyses on the structural effects of scaffolding on different phases of the CPS process were lacking. This may partly explain the current lack of empirical research on how scaffolds should be designed and provided at specific instances during CPS to effectively address students' learning needs (e.g., \cite{nasir_2023HMMbased}). The problem solving process taken in domain-specific CPS tasks can be described in phases using a theoretical model relevant to its subject domain (e.g., mathematics \cite{rott_2021Descriptive}). By adopting such models in the learning design of the CPS task, a structured approach can be taken to analyse the structural effects of scaffolding. In this study, we apply two analytical methodologies to examine aggregated measures and the temporality of CPS behavioural indicators within each phase, as well as across various phases in the CPS process. Our goal is not merely to observe differences between the two scaffolding conditions, but also to examine them through two analytic approaches in order to highlight their relative strengths and weaknesses.

\section{Related Work}
Previous research that applied scaffolding to support learners during CPS largely examined its effects on students' learning outcomes and collaborative behaviours.

\subsection{Effects of scaffolding on learning in CPS}
During collaboration, students can learn to acquire domain-specific knowledge \cite{teasley_1995role,baker_2009Argumentative} or domain-general behaviours \cite{rogat_2015Interrelation} through various approaches and support. In particular, scaffolding through the use of scripting has been widely shown to be beneficial to learning. For example, King et al. \cite{king1991effects} demonstrated how scripting, through the use of the Guided Strategic Problem Solving (GSPS) procedure, enabled students to achieve better performance and was effective in encouraging student dyads to request and provide elaboration during problem solving. The use of collaborative scripts, which has been explored across numerous studies within the CSCL community, was shown in a meta-analysis to have a large positive effect on the acquisition of collaboration skills ($d = 0.95$). Yet, there are various concerns about its applications. One concern is that providing an extensive amount of instruction in scripting (overscripting) \cite{dillenbourg_2002Overscripting} could increase cognitive load and decrease student autonomy, consequently hurting student motivation. This has led to studies examining the optimal amount of scaffolding required to support student learning.

There remains a mixed consensus on how best to scaffold learners through the use of scripting. Some studies argue for a structured script that provides instruction to students throughout the entire process of learning complex skills \cite{Kirschner_2006}. In contrast, others contend for the use of minimal scripting structures \cite{duffy_2013Constructivism}. A more nuanced approach involving scripting that is customised to learners' needs has also been advocated in various studies \cite{cai_2024efficiency}. Although these studies have elaborated extensively on the overall impact of such scaffolds on learning outcomes, collaborative quality, and even the psychological needs of students, few studies that applied scaffolding in CPS \cite{li_2025Effect} have investigated these impacts across different phases of the problem-solving process. In this study, we contribute to advancing this line of investigation by analysing how two scaffolding conditions (i.e., maximal and minimal) shape students' engagement and behaviours across CPS phases, to inform future scaffold design and teaching practices of CPS processes.

\subsection{Analysis of CPS processes}
During CPS, group members engage with one another adaptively to produce dynamic processes involving the perspectives of both individuals and groups \cite{hesse_2015Framework,luckin_2017Solved}. Existing studies have sought to understand these complexities by investigating CPS processes through the use of analytical approaches and data mining techniques. For example, Ouyang et al.\cite{ouyang_2023artificial} applied Epistemic Network Analysis (ENA) and Hidden Markov Model (HMM) using multimodal data to examine the structural and temporal characteristics of CPS respectively. Using ENA, the connection structure and association of coded collaborative indicators were visualised as a network to derive insights into the different clusters obtained through hierarchical clustering. The application of HMM produced hidden states, and the probabilities of observed states within each of these hidden states described the transition of coded collaborative indicators. In another study, Taylor et al. \cite{taylor_2024Quantifying} performed Sequential Pattern Mining (SPM) on CPS subskills to identify recurring patterns that were representative of CPS processes. These emergent patterns were associated with performance outcomes to determine approaches that could enhance CPS processes.

While these studies presented meaningful approaches towards uncovering processes in CPS, the analyses were performed at the group level across the whole CPS task. Although insights into the overall process of CPS were obtained, they do not capture the nuances associated with individual students and their respective developments during the problem solving process in CPS. Students can progress through different phases during problem solving \cite{polya_1957How}, and individual behaviour can impact group-level interaction \cite{luckin_2017Solved}. Hence, our study analyses students' engagement and behaviours at the individual level in each distinct problem solving phase of CPS. Furthermore, we also expand the range of methodological approaches that can be used to analyse CPS processes by exploring the value of Heterogeneous Interaction Network Analysis (HINA) \cite{feng2025hina}. HINA provides a methodological framework for modelling and examining learning processes using heterogeneous interaction networks (HINs) \cite{feng2025hina}. HINs consist of multiple node types, with edges connecting nodes from different sets, thereby capturing complex multi-entity interactions inherent in learning environments \cite{feng2024heterogenous,feng2025analyzing}. A common form of HIN, the bipartite network, is particularly suitable for representing relationships between two distinct node sets. Currently, the use of HINA has not been applied in studies involving CPS. Specifically, we apply HINA to model and analyse multi-entity relationships in CPS processes and quantify the structural characteristics of CPS processes. Through this, we consider its complementary value to other approaches, in particular to SPM.

\section{Current Study}
This study explores the application of HINA and SPM to investigate the effects of minimal and maximal scaffolding on patterns and associations of individual CPS behaviours. While the minimal and minimal scaffolds were both designed to consider essential principles of collaborative and problem solving scripting \cite{fischer_2013Script}, the maximal scaffold was designed with additional guidance and prompts (further detailed in section \ref{sec:min} and \ref{sec:max}). The method and approach used in this study present key contributions to the literature by analysing the distribution of demonstrated CPS indicators across and within different CPS phases to study the effects of scaffolding. We address two research questions in this study:
\newline \textbf{RQ1}: How does individual dialogic engagement across CPS phases differ between minimal and maximal scaffolding?\newline
\textbf{RQ2}: How do CPS behaviours in different CPS phases differ between minimal and maximal scaffolding?

\section{Methodology}
\subsection{Context and Participants}
The participants involved in the study consisted of 78 Secondary-level students aged 14 - 16 years old from a public school. A pre-test was administered to assess their prerequisite domain knowledge and ability to solve a similar problem-solving question as the CPS task. Participants within the same mathematics class were grouped into triads, ensuring there were no significant differences in pre-test scores between the triads. This was done to form groups of similar knowledge and status \cite{Dillenbourg1999}. A total of 26 triads (18 mixed-sex, 4 all-female, 4 all-male) were formed and randomly allocated to the  minimal scaffold condition ($n = 13$ triads) or the maximal scaffold condition ($n = 13$ triads) 

In both conditions, students were given the same mathematics problem-solving question that involved applying the Pythagorean theorem and quadratic equations. The question was provided at the start of the CPS task, and students had to follow a structured problem-solving process \cite{polya_1957How} to solve the question and complete the CPS task. The task was hosted on an online learning platform accessible to all students. The problem-solving process was structured into four CPS phases: \textit{problem identification} (A1), \textit{ideation, planning and decision making} (A2), \textit{plan implementation and solution generation} (A3), and \textit{solution checking, problem extension and reflection} (A4). Students could only progress to the next phase after working on the instructions given in the previous phase. All students were given 50 minutes to work on the CPS task with their assigned triad. Only the amount of scripting structured into the CPS task was varied between the different scaffold conditions. 

\subsubsection{Minimal scaffold condition} \label{sec:min}
In each CPS phase, students were given contextualised hints in the form of questions to prompt their thinking and consideration of particular aspects of the problem-solving question. Individual and transactive activities (e.g., crafting individual contributions, group discussion, and consolidation of group discourse) were also included to ensure that students continued to collaborate while also working on the scaffolds in each phase individually. This also sought to reduce the collaborative cognitive load among students \cite{kirschner_2018Cognitive}.

\subsubsection{Maximal scaffold condition} \label{sec:max}
In addition to the scaffold provided in the minimal condition, the maximal scaffold condition also provided students with principles and pre-populated prompts that sought to guide students' demonstration of problem-solving skills \cite{schoenfeld_1980Teaching} in each CPS phase. Furthermore, students were asked to select a dialogue act from a list and apply it during the transactive activities where they either responded to their group or a group member \cite{mercer_2000Words,michaels_2008Deliberative}.

\subsection{Data Collection and Processing}
\begin{figure}[!ht]
\centering
\includegraphics[clip, trim=0cm 0cm 0cm 0cm, width=\textwidth]{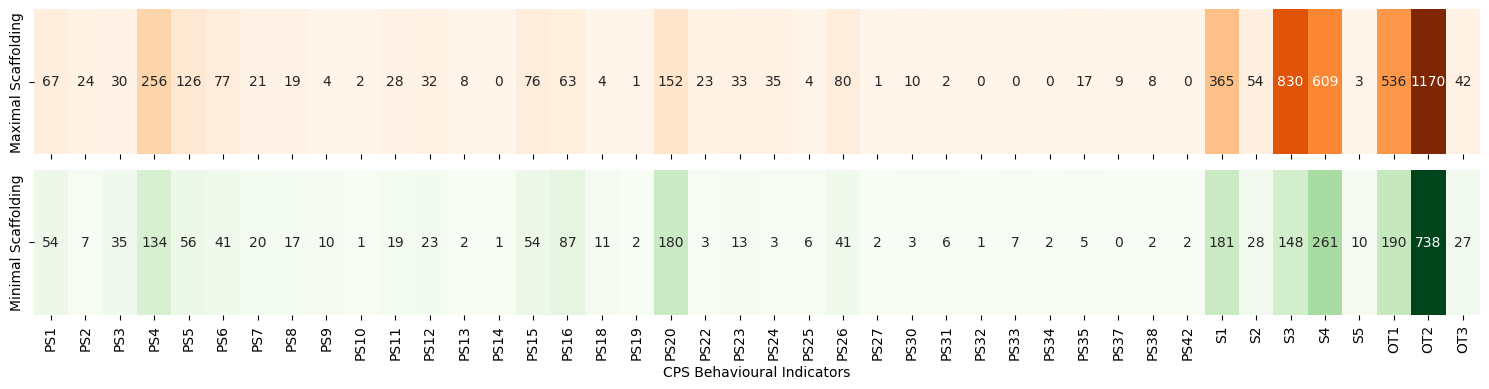}
\caption{Distribution of coded behavioural indicators in the dataset for each scaffold condition.}
\label{fig:data}
\end{figure}

Transcription of dialogue data and log data on students' progression to each CPS phase was collected for each student. The timestamps for these two data sources were matched to determine the conversation of students in each phase. Transcription utterances were coded at the indicator level using a theoretical framework that involves behavioural indicators of problem-solving (\textit{PS1} - \textit{PS42}) and scripting (\textit{S1} - \textit{S5}) \cite{wong_2025Rethinking}. These indicators corresponded to relevant subskills of CPS processes. As students might choose to engage in processes that are unrelated to problem solving or scripting \cite{skinner_2009Motivational}, three additional indicators were included into the framework: "technical issues or logistical tasks related to the learning environment" (\textit{OT1}), "socialising" (\textit{OT2}), and "refocusing to disrupt engagement in technical issues, logistical tasks or socialising" (\textit{OT3}). The theoretical framework of these indicators is provided in Table \ref{tab:cpsframework}. Utterances were coded across 24 distinct indicators, with a Cohen's kappa interrater reliability of $0.847$ on $17.5\%$ of the total coded utterances. Figure \ref{fig:data} shows the distribution of coded behavioural indicators across the maximal ($n = 4821$) and minimal ($n = 2433$) scaffold conditions, with a heatmap distinguishing indicators that were in higher and lower amounts in the dataset.

\begin{table}[!ht]
\centering    
\resizebox{\textwidth}{!}{
\begin{tabular}{ll|c|p{13.5cm}}
\hline
\multicolumn{1}{l|}{\textbf{Dimension}} & \textbf{Subskill} & \textbf{Label} & \textbf{Indicator} \\ \hline
\multicolumn{1}{l|}{Problem Solving} & SS1: Sense-making & PS01 & Talking about the task questions in general terms to understand about the problem-solving task \\
\multicolumn{1}{l|}{} & & PS02 & Explaining ideas or concepts in the problem-solving task with reference to prior knowledge or definitions from information sources \\
\multicolumn{1}{l|}{} & & PS03 & Addressing difficulties or limitations that obstruct problem solving \\ \cline{2-4}
\multicolumn{1}{l|}{} & SS2: Building shared understanding & PS04 & Asking questions to clarify understanding, ideas or contributions \\
\multicolumn{1}{l|}{} & & PS05 & Answering questions to clarify understanding, ideas or contributions \\
\multicolumn{1}{l|}{} & & PS06 & Reiterating or paraphrasing oneself or others’ ideas or contributions \\
\multicolumn{1}{l|}{} & & PS07 & Adapting and building on the ideas or contributions of others \\
\multicolumn{1}{l|}{} & & PS08 & Stating agreement with others \\
\multicolumn{1}{l|}{} & & PS09 & Discovering perspectives and abilities of group members \\
\multicolumn{1}{l|}{} & & PS10 & Sharing information from sources which contribute to formulating the problem-solving task \\
\multicolumn{1}{l|}{} & & PS11 & Stating disagreement with others \\
\multicolumn{1}{l|}{} & & PS12 & Constructing arguments in favour of one's own ideas or contributions \\
\multicolumn{1}{l|}{} & & PS13 & Resolving differences \\
\multicolumn{1}{l|}{} & & PS14 & Reaching a compromise with others \\
\multicolumn{1}{l|}{} & & PS15 & Identifying and abstracting relevant information about the task context \\
\multicolumn{1}{l|}{} & & PS16 & Establishing connections and patterns between relevant information in the problem-solving task \\
\multicolumn{1}{l|}{} & & PS17 & Dissecting the problem into smaller tasks \\ \cline{2-4}
\multicolumn{1}{l|}{} & SS3: Formulating a solution & PS18 & Building a representation of the problem-solving task \\
\multicolumn{1}{l|}{} & & PS19 & Creating an ordered step-by-step plan \\
\multicolumn{1}{l|}{} & & PS20 & Proposing ideas or specific solution methods to solve the task questions \\ \cline{2-4}
\multicolumn{1}{l|}{} & SS4: Defining roles and responsibilities & PS21 & Discussing required roles and collaborative interaction to address the problem-solving task \\
\multicolumn{1}{l|}{} & & PS22 & Coordinating sub-tasks to be performed \\ \cline{2-4}
\multicolumn{1}{l|}{} & SS5: Reaching a solution & PS23 & Sharing contributions and findings of individual and group sub-tasks \\
\multicolumn{1}{l|}{} & & PS24 & Providing an answer to the task questions \\
\multicolumn{1}{l|}{} & & PS25 & Responding to or acknowledging the contributions of others \\ \cline{2-4}
\multicolumn{1}{l|}{} & SS6: Maintaining roles and responsibilities & PS26 & Discussing the progress and status of individual and group sub-tasks \\
\multicolumn{1}{l|}{} & & PS27 & Providing feedback on the progress and status of individual or group sub-tasks \\
\multicolumn{1}{l|}{} & & PS28 & Recognising strengths and weaknesses of self and others \\
\multicolumn{1}{l|}{} & & PS29 & Adapting group organisation to adjust individual and group sub-tasks \\ \cline{2-4}
\multicolumn{1}{l|}{} & SS7: Maintaining shared understanding & PS30 & Providing feedback or instructional support to others \\
\multicolumn{1}{l|}{} & & PS31 & Using feedback provided to clarify or elaborate own ideas \\
\multicolumn{1}{l|}{} & & PS32 & Making iterative adaptations to the plan based on outcomes, new information and new ideas \\ \cline{2-4}
\multicolumn{1}{l|}{} & SS8: Evaluating the solution & PS33 & Anticipating issues or errors \\
\multicolumn{1}{l|}{} & & PS34 & Testing to detect working order \\
\multicolumn{1}{l|}{} & & PS35 & Detecting and hypothesising issues or errors \\
\multicolumn{1}{l|}{} & & PS36 & Identifying the need for additional information, resources or tasks to address issues or fix errors \\
\multicolumn{1}{l|}{} & & PS37 & Addressing issues or fixing errors \\
\multicolumn{1}{l|}{} & & PS38 & Agreeing the sub-goals or goal-state have been effectively solved to answer the problem \\ \cline{2-4}
\multicolumn{1}{l|}{} & SS9: Reflecting & PS39 & Reusing, remixing, and integrating ideas to develop alternative strategies for flawed solutions \\
\multicolumn{1}{l|}{} & & PS40 & Building on others’ ideas to improve alternative strategies \\
\multicolumn{1}{l|}{} & & PS41 & Discussing the limitations of the current solution for future problem-solving tasks \\ \cline{2-4}
\multicolumn{1}{l|}{} & SS10: Evaluating on group work & PS42 & Discussing group dynamics, effort, strengths and weakness \\ \hline
\multicolumn{1}{l|}{Scripting} & SC11: Using scripting & S1 & Discussing understanding of script components \\
\multicolumn{1}{l|}{} & & S2 & Prompting responses or actions from others to script components \\
\multicolumn{1}{l|}{} & & S3 & Responding to script components \\ \cline{2-4}
\multicolumn{1}{l|}{} & SC12: Regulating scripting & S4 & Discussing work status and progress on script components \\
\multicolumn{1}{l|}{} & & S5 & Disconnecting with group progress and usage of script components \\ \cline{2-4}
\hline
\multicolumn{2}{l|}{\textbf{Process not related to Problem Solving or Scripting}} & \textbf{Label} & \textbf{Indicator} \\ \hline
\multicolumn{2}{l|}{Other: Other engagements during task} & OT1 & Technical issues or logistical tasks related to the learning environment \\
\multicolumn{1}{l}{} & & OT2 & Socialising \\
\multicolumn{1}{l}{} & & OT3 & Refocusing to disrupt engagement in technical issues, logistical tasks or socialising \\ \hline
\end{tabular}
}
\caption{Dimensions, subskills, and indicators of the CPS framework.}
\label{tab:cpsframework}
\end{table}

\subsection {Analysing Individual Dialogic Engagement and Behaviours in CPS using HINA (RQ1 and RQ2)}
In this study, we constructed two HINs to model individual dialogic engagement and behaviours in CPS and used the individual-level and dyadic-level functions of the \texttt{HINA web tool} \footnote{\scriptsize{\url{http://hina-network.com}}} to analyse these networks. The first HIN involved a student–phase network $H_{S,P}$ that captured temporal associations between students ($S$) and CPS phases ($P$) under the two scaffolding conditions (\textbf{RQ1}). Formally, we constructed a heterogeneous interaction network (HIN) $H_{S,P} = (S, P, E, w)$, where $S = \{s_1, s_2, \dots, s_N\}$ is the set of students, $P = \{\text{A1}, \text{A2}, \text{A3}, \text{A4}\}$ is the set of four CPS phases, $E \subseteq S \times P$ is the set of edges connecting students to phases, and $w:E_{S,P} \rightarrow Z^{+}$ is a weight function where $w(s, p)$ denotes the frequency of co-occurrences between student $s \in S$ and phase $p \in P$.

Each edge $(s, p) \in E_{S,P}$ represents a temporal co-occurrence, capturing how frequently a student engaged in dialogue within a specific CPS phase. The weights $w(s, p)$ aggregate these interactions over the entire problem-solving cycle, providing a basis for deriving the total participation of individual dialogic engagement (quantity) and the variety of phase involvement (diversity). The quantity ($Q_i$) measures the volume of a student's total participation across CPS phases. A student who interacts frequently throughout the CPS phases will have a high quantity score, reflecting their overall level of activity. The diversity ($D_i$) measures the distribution of a student's engagement across the different CPS phases. It is calculated using entropy, quantifying how evenly a student's interactions are spread across the four phases (A1 - A4) \cite{feng_2020Mixing}. A higher diversity score indicates that a student is more broadly engaged in all aspects of the CPS process. In contrast, a lower score suggests that engagements were more concentrated on only one or two phases.

This two-dimensional perspective is crucial, as it enables a nuanced differentiation between engagement strategies that would not be achieved by a single metric. For example, it distinguishes a student who participated extensively in a single phase from one who contributed more moderately but distributed their engagements evenly across all four CPS phases, revealing fundamentally different approaches to the collaborative task. The quantitative nature of these metrics enables direct statistical comparison of engagement patterns across different scaffolding conditions, revealing how instructional scaffolding influenced student engagement differently during CPS. As such, the quantity and diversity of individual student engagement across the CPS phases were compared between maximum and minimum scaffolding conditions using the Mann-Whitney $U$ test \cite{feng2025analyzing}. 

The other HIN involved a behaviour–phase network $H_{B,P}$ that captured associations between behavioural indicators ($B$) and CPS phases ($P$) for each scaffolding condition (\textbf{RQ2}). Statistically significant edges were identified to reveal meaningful associations between behavioural indicators and CPS phases. The network is defined as $H_{B,P} = (B, P, E, w)$, where $B = \{ PS01, PS02, PS03, \ldots \}$ is the set of fine-grained behavioural indicators, $P = \{ \text{A1}, \text{A2}, \text{A3}, \text{A4} \}$ is the set of four CPS phases, $E_{B,P} \subseteq B \times P$ is the set of edges connecting behaviours to phases, and $w:E_{B,P} \rightarrow Z^{+}$ is a weight function where $w(b, p)$ counts the co-occurrence frequency of behaviour $b$ and phase $p$, such that $b \in B$ and $p \in P$.

The $H_{B,P}$ models how specific learning behaviours manifest across different phases of collaborative problem-solving in each scaffolding condition. Rather than merely counting behaviour frequencies, the pruned $H_{B,P}$ captures which behavioural patterns are \emph{characteristic} of each CPS phase (e.g., whether "asking questions to clarify understanding" was prominent during the initial design phase or the final refinement phase). This allows us to move beyond \emph{whether} students are engaged to understand \emph{how} they are engaged at different points in the collaborative learning processes. 

The dyadic-level function \cite{feng2024heterogenous} within the HINA framework was used to identify statistically significant behaviour-phase associations, distinguishing meaningful engagement patterns from random co-occurrences. For each edge $(b, p) \in E_{B,P}$, we tested whether its observed weight $w(b, p)$ represented a meaningful interaction beyond what would occur by chance. This was achieved by comparing $w(b, p)$ against a null model where edges were randomized without preserving node degrees, following a binomial distribution. Edges with weights exceeding the 95\textsuperscript{th} percentile of the null distribution ($p < 0.05$) were retained as statistically significant, forming a pruned network of robust behaviour-phase associations for each scaffolding condition. The visualisation of the pruned $H_{B,P}$ networks for both the maximum and minimum scaffolding conditions was produced from the \texttt{HINA web tool}. The resulting significant edges reveal which specific behaviours are aligned with particular CPS phases in different scaffolding conditions. For example, a significant connection between behavioural indicators such as \textit{PS18} - \textit{PS20} (i.e., `\textit{formulating a solution}' subskill) and the A2 phase (\textit{ideation, planning and decision making}) would validate expected theoretical relationships, while the significant connection between \textit{PS01} (i.e., `\textit{sense-making}' subskill) and the A4 phase (\textit{solution checking, problem extension and reflection}) in the maximal scaffolding condition reveals the observation of students reverting to behaviours associated with earlier phases in the problem solving process despite them reaching to later CPS phases in their progress. This method thus provides both confirmatory and exploratory insights into the temporal dynamics of collaborative learning, highlighting how specific scaffolding conditions shape behavioural pathways through problem-solving activities.

\subsection{Analysing Patterns across Individual Sequences of CPS Behaviours using SPM Analysis (RQ2)}
SPM involves finding the set of subsequences that are frequent in one sequence or a set of sequences \cite{HAN2012585}. In particular, SPM is applied to derive frequently occurring sequences from the collection of individual students' sequences of behavioural indicators during CPS. The sequence of behavioural indicators during CPS phases for students is defined as $S_{p, i} = \langle b_1b_2 \ldots b_j\rangle$, where $1 \leq i \leq n_p$ such that $n_p$ is the number of students who demonstrated behavioural indicators for a given CPS phase, and $b_1, b_2, \ldots , b_j \in B$ such that $j$ is the number of behavioural indicators in the sequence. The collection of student sequences for a given CPS phase is defined as $SEQ_{p} = \langle S_{p,1}, S_{p,2}, \ldots ,S_{p,n_p} \rangle$. A sequence $\alpha_p=\langle a_1a_2 \ldots a_j \ldots a_{m_p}\rangle$ is a subsequence $SEQ_{p}$ if there exist integers $1 \leq j_1 <j_2 < \ldots <j_{m_p} \leq n_p$ such that $a_1 \subseteq S_{j_{p,1}}, a_2 \subseteq S_{j_{p,2}},\ldots , a_{m_p} \subseteq S_{j_{p,n_p}}$. For example, if $SEQ_{A1} = \langle \{PS01, PS04, PS05 \}, \{PS04, PS04 \}, \{PS01, PS05, PS06 \} \rangle$ and $\beta_{A1}=\langle \{PS01\}, \{PS01, PS06 \}\rangle$, then $\beta_{A1}$ is a subsequence of $SEQ_{A1}$.

The \texttt{PrefixSpan} algorithm \cite{jianpei_2001PrefixSpan}, which uses a pattern-growth method to determine these subsequences, is performed on $SEQ_p$. The algorithm progresses until the
subsequences $\alpha_p$ whose occurrence frequency in $SEQ_p$ is no less than the user-specified frequency threshold (\textit{min\_support}). The algorithm was implemented in \texttt{PySpark} \footnote{\scriptsize{\url{https://spark.apache.org/docs/latest/api/python/reference/api/pyspark.ml.fpm.PrefixSpan.html}}}, where providing a \textit{min\_support} of 0.3 would yield subsequences that are frequently occurring in at least 30\% of $n_P$. In this study, the \textit{min\_support} values selected to obtain the subsequences for the different CPS phases were chosen to obtain at least one sequence that had three behavioural indicators to avoid having an excessive number of sporadic patterns \cite{fournier2017survey}. Furthermore, consecutive occurrences of the same behavioural indicator within each student's sequence were merged as a single indicator to minimise the noise in identifying frequently occurring patterns introduced by fine-grained utterance-level coding (i.e., a continuous statement made by a student could be split across multiple utterances that are coded on the same indicator). Only subsequences with at least two behavioural indicators were included for analysis since the goal is to examine the progression from one behavioural indicator to another. A search is then performed to count the number of times that a subsequence appears in the sequence of students for each scaffolding condition. The subsequences obtained for each CPS phase are visualised as a flow diagram, showing also the percentage of students with the subsequence and its percentage contribution to total subsequence occurrences. 

\section{Results}
\subsection{Differences in Individual Dialogic Engagement between Scaffolds (RQ1)}
\begin{figure}[!ht]
\centering
\includegraphics[clip, trim=0cm 0cm 0cm 0cm]{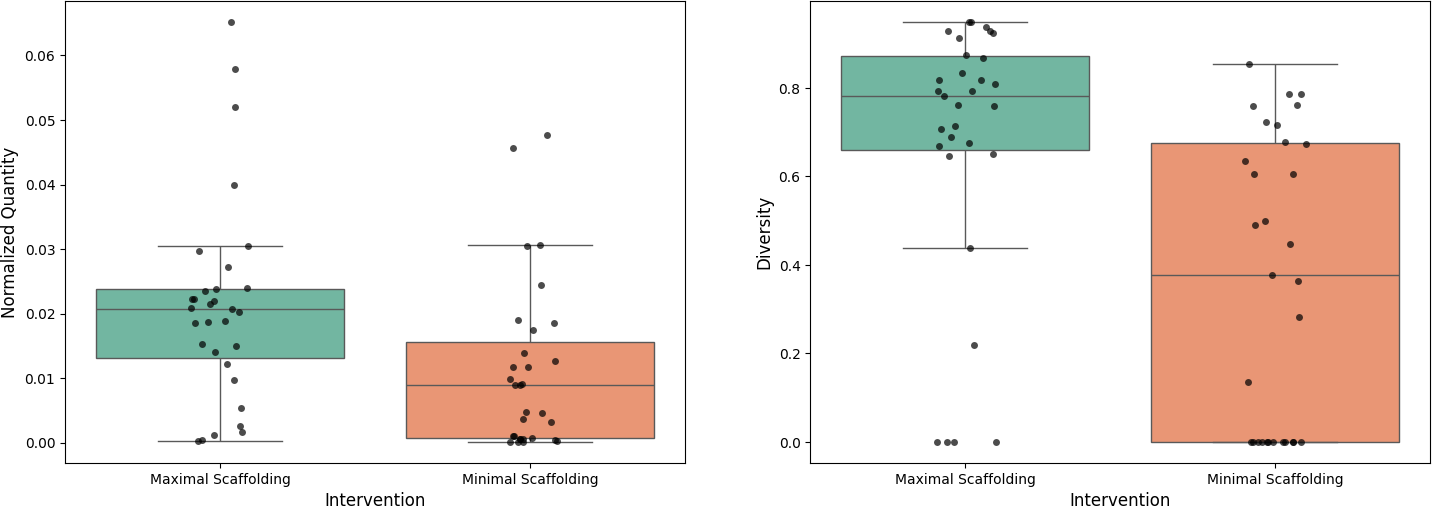}
\caption{Boxplot comparing the normalised quantity (left) and diversity (right) of individual dialogic engagement across CPS phases between maximal and minimal scaffolding.}
\label{fig:boxplot}
\end{figure}

Figure \ref{fig:boxplot} summarises the CPS indicator quantity and diversity-based differences between the minimal and maximal scaffolding settings. It is observed in Figure \ref{fig:boxplot} that the median value for normalised quantity and diversity for the maximal scaffolding condition was higher than that for the minimal scaffolding condition. The range of the normalised quantity and diversity for the minimal scaffolding condition was wider compared to the maximal scaffolding condition, with fewer outliers. In particular, student outliers in the maximal scaffolding condition were observed to be above the median for normalised quantity but below the median for diversity. This shows that the quantity and diversity of indicators observed in the maximal setting were higher than in the minimal setting, with smaller variance in numbers. 

Further analysis showed that the difference in normalised quantity of dialogic engagements of individual students across CPS phases between maximal and minimal scaffolding conditions was statistically significant ($U = 701, p = 0.00195, RBC = 0.459$), with a close to medium effect size \cite{cohen_1992power}. The difference in diversity of dialogic engagements of individual students across CPS phases between maximal and minimal scaffolding conditions was also statistically significant ($U = 755, p = 0.0001, RBC = 0.571$) with a medium effect size. These observations show that the maximal scaffold enabled students to be more participatory and diverse in their engagement across the CPS phases compared to minimal scaffolding. The analysis of individual dialogic engagement also allowed for the identification of students who fell below the median value of diversity. These students might not be able to engage in the same way as their peers across the CPS phases, despite having the maximal scaffolding. However, we did not observe outliers in the analysis for diversity in the minimal scaffold condition. Some particular students with the maximal scaffold were increasingly unwilling to engage across all CPS phases, possibly due to the increased number of choices (e.g., selecting a dialogue act for collaboration with group members) that came with the additional guidance and instructions \cite{katz_2007When}. 

\subsection{Differences in CPS Behaviours between Scaffolds (RQ2)}
\subsubsection{HINA}
\begin{figure}[!ht]
\centering
\includegraphics[clip, trim=0cm 0cm 0cm 0cm, width=\textwidth]{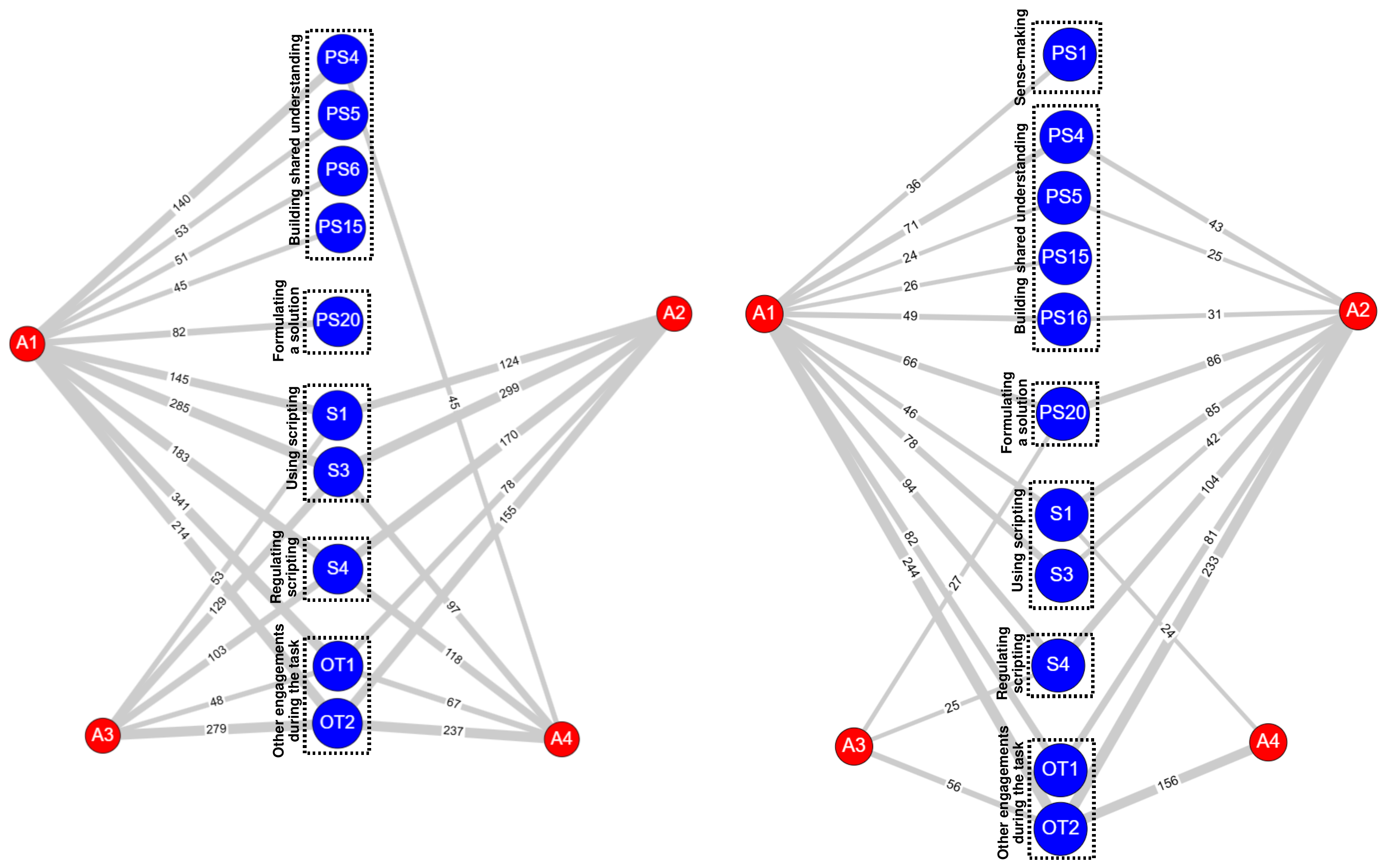}
\caption{Pruned behaviour-phase network with statistically significant edges for the maximal scaffold (left) and minimal (right) condition.}
\label{fig:hina_behavioural}
\end{figure}

The visualisation of the pruned behaviour-phase network ($H_{B,P}$) is provided in Figure \ref{fig:hina_behavioural} for the two scaffold conditions. Overall, students in both maximal and minimal scaffold conditions engaged in similar subskills in the A1 CPS phase (i.e., building shared understanding, formulating a solution, using scripting, and regulating scripting). However, it is observed that students demonstrated the `\textit{sense-making}' subskill in the minimal scaffold (i.e., \textit{PS1} indicator). The increased amount of instructions and guidance provided in the maximal scaffold condition could have led to students' over-reliance on these resources to understand the task, rather than working with their group members to make sense of the given problem. Yet, the association of the \textit{PS1} indicator to the A1 CPS phase ($\text{edge weight}= 36$) was much lower compared to the association of the \textit{OT2} indicator with the A1 CPS phase ($\text{edge weight}= 244$) in the minimal scaffold condition. Although the association of the \textit{OT2} indicator with the A1 CPS phase in the maximal scaffold condition was of similar value to that in the minimal scaffold condition ($\text{edge weight}= 214$), the association of the other problem solving and scripting indicators to the A1 CPS phase was much higher in the maximal scaffold condition than in the minimal scaffold condition. Even though the maximal scaffold provided excessive guidance in the A1 CPS phase and moved students away from engaging in sense-making, it reinforced students' problem solving and scripting behaviours in this initial phase of CPS.

Students in this study were unfamiliar with the scaffold provided in the maximal scaffold and spent a substantial amount of effort working with it. As such, we observed a high association between scripting indicators (i.e., \textit{S1}, \textit{S3} and \textit{S4}) across all CPS phases in the maximal scaffold. However, the association between scripting indicators across all CPS phases in the minimal scaffold was much lower, with no association of indicators related to using scripting in the A3 and A4 CPS phases. Students were evidently using and regulating the use of scripting more often in the maximal scaffold condition than in the minimal scaffold condition. These students demonstrated an interest and willingness to accept and act upon the additional instructions provided in the maximal scaffold. However, this raises the concern that students may develop an over-reliance on the maximal scaffold, especially since they were extensively using and regulating scripting across all CPS phases. Scaffolding should provide initial assistance to learners and be progressively faded out to increase learner autonomy and independence when solving problems \cite{Pea01072004}.

It was interesting to note that there were no associations between problem-solving indicators and the A2, A3, and A4 CPS phases in the maximal scaffold condition. Students instead sought to build shared understanding and move toward formulating a solution in the A1 CPS phase, even though the provided instructions and scaffolds were only related to \textit{problem identification}. While having a determined order of problem solving structured into the CPS task is helpful for secondary school learners to understand what problem solving entails, this observation of students not following what was designed in the CPS task aligns with other studies that described real-world problem solving to be a non-linear, subtle and unpredictable process \cite{rott_2021Descriptive}. In the maximal scaffold condition, students immediately proceeded to formulate a solution upon building shared understanding in the A1 CPS phase. They had sufficient clarity on what they wanted to work on, and hence, in A2, A3 and A4 CPS phase, they were using what they had decided upon in A1 to follow the scripting given to complete the CPS task. Of particular concern is that, although students with the minimal scaffold similarly sought to formulate a solution (i.e., \textit{PS20} indicator) in the A1 CPS phase as students with the maximal scaffold, these students continued to demonstrate this behaviour in the A2 phase (i.e., `\textit{ideation, planning and decision making}')  and also in the A3 phase (i.e., `\textit{plan implementation and solution generation}'). Considering that students would seek to exert their own expression of the problem solving process, it is apparent that the provision of substantive instructions and inclusion of relevant collaborative activities could potentially prevent them from being stuck on particular problem solving behaviours. 

It is only in the A4 CPS phase that we observed students in the maximal scaffold condition once again demonstrate a problem solving indicator (i.e., \textit{PS5}: answering questions to clarify understanding, ideas or contributions). However, in the minimal scaffold condition, we no longer observed an association between problem-solving indicators and the A4 CPS phase. In this phase where students were supposed to engage in `\textit{solution checking, problem extension and reflection}’, there was an absence of problem solving indicators relating to `\textit{evaluating the solution}', `\textit{reflecting}' and `\textit{evaluating teamwork}' subskills in both the maximal and minimal scaffold conditions. The current design of these scaffolds appears to be insufficient to induce these behaviours among students. Further refinement of the scaffolds would be required to support metacognitive behaviours in the CPS process to support improvements in problem solving abilities \cite{stillman_2010Metacognition}.

\subsubsection{SPM Analysis}
\begin{figure}[!ht]
\centering
\includegraphics[clip, trim=0cm 4cm 0cm 4.1cm, width=\textwidth]{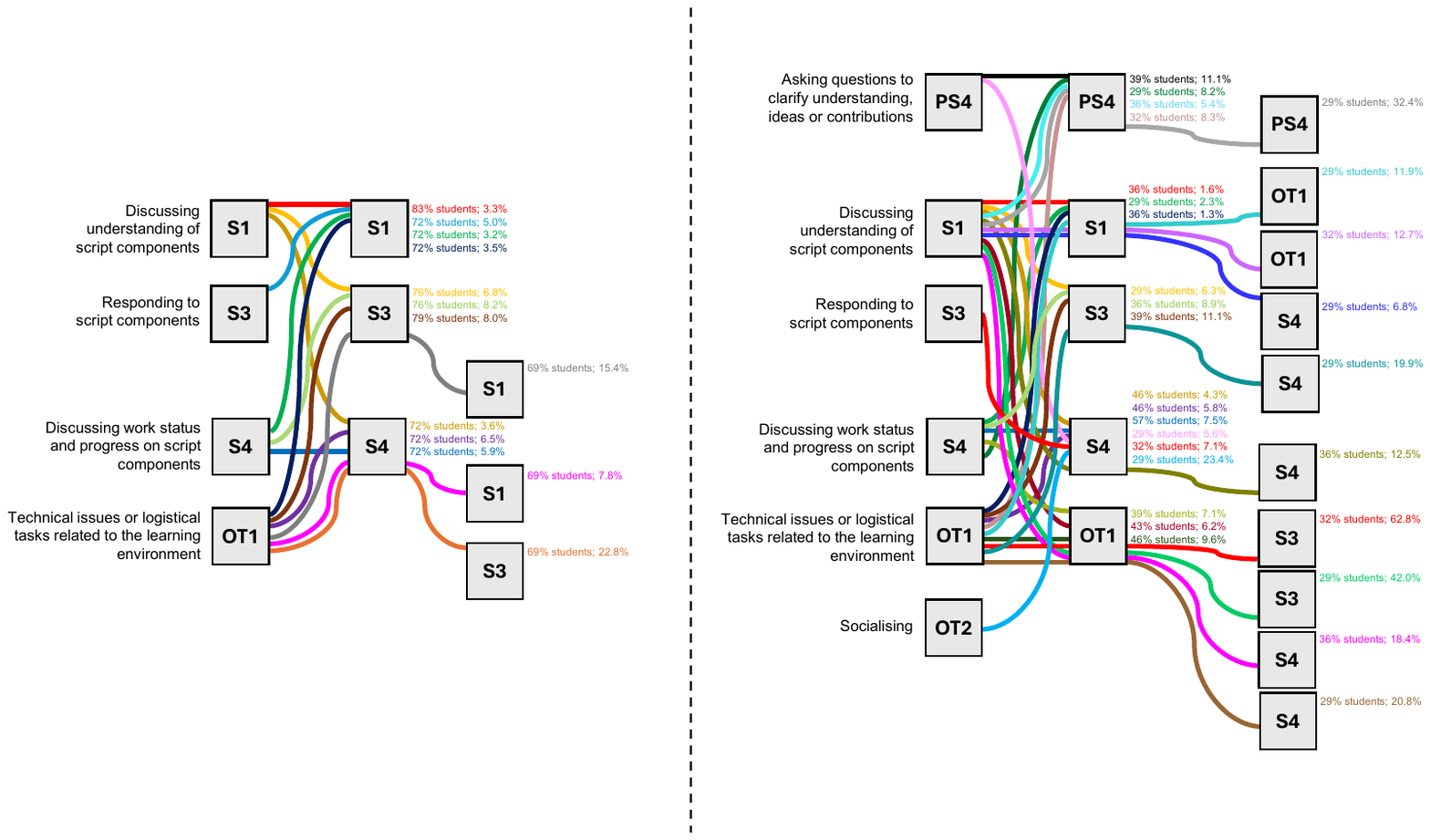}
\caption{Frequently occurring sequence of CPS indicators in maximal scaffolding among 29 students (left) compared to that in minimal scaffolding among 28 students (right) in the \textit{problem identification} phase.}
\label{fig:spm_segment1}
\end{figure}

\begin{figure}[!ht]
\centering
\includegraphics[clip, trim=0cm 4.6cm 0cm 5.6cm, width=\textwidth]{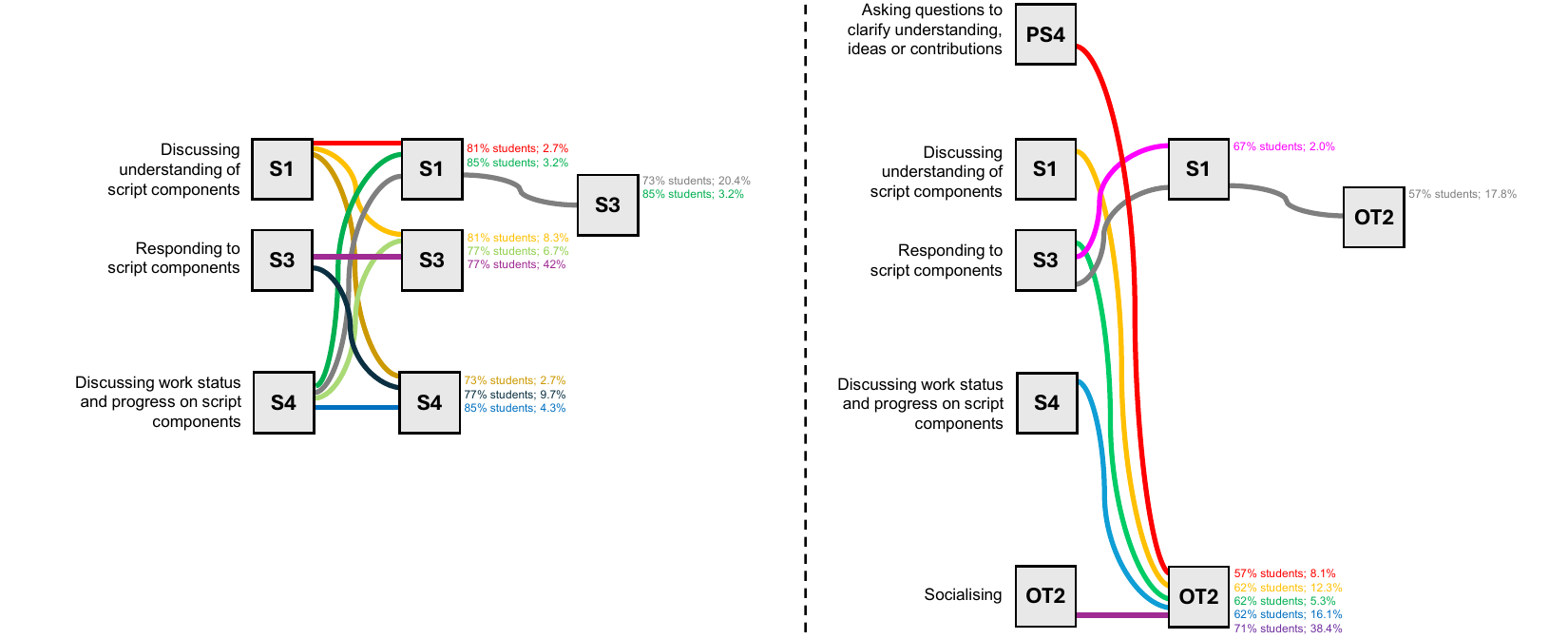}
\caption{Frequently occurring sequence of CPS indicators in maximal scaffolding among 26 students (left) compared to that in minimal scaffolding among 21 students (right) in the \textit{ideation, planning and decision making} phase.}
\label{fig:spm_segment2}
\end{figure}

\begin{figure}[!ht]
\centering
\includegraphics[clip, trim=0cm 3.5cm 0cm 3.2cm, width=\textwidth]{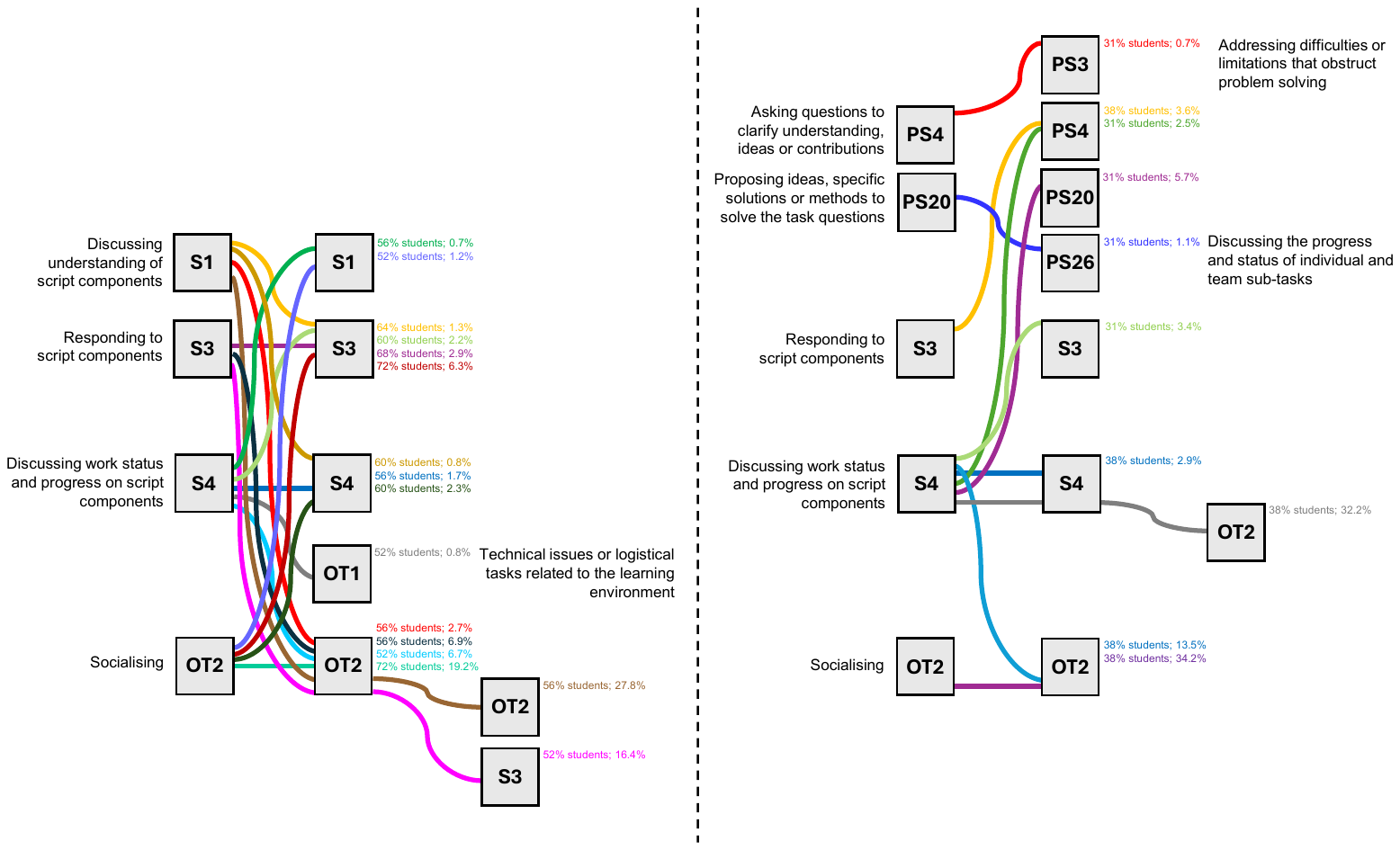}
\caption{Frequently occurring sequence of CPS indicators in maximal scaffolding among 25 students (left) compared to that in minimal scaffolding among 13 students (right) in the \textit{plan implementation and solution generation} phase.}
\label{fig:spm_segment3}
\end{figure}

\begin{figure}[!ht]
\centering
\includegraphics[clip, trim=0cm 5.5cm 0cm 5cm, width=\textwidth]{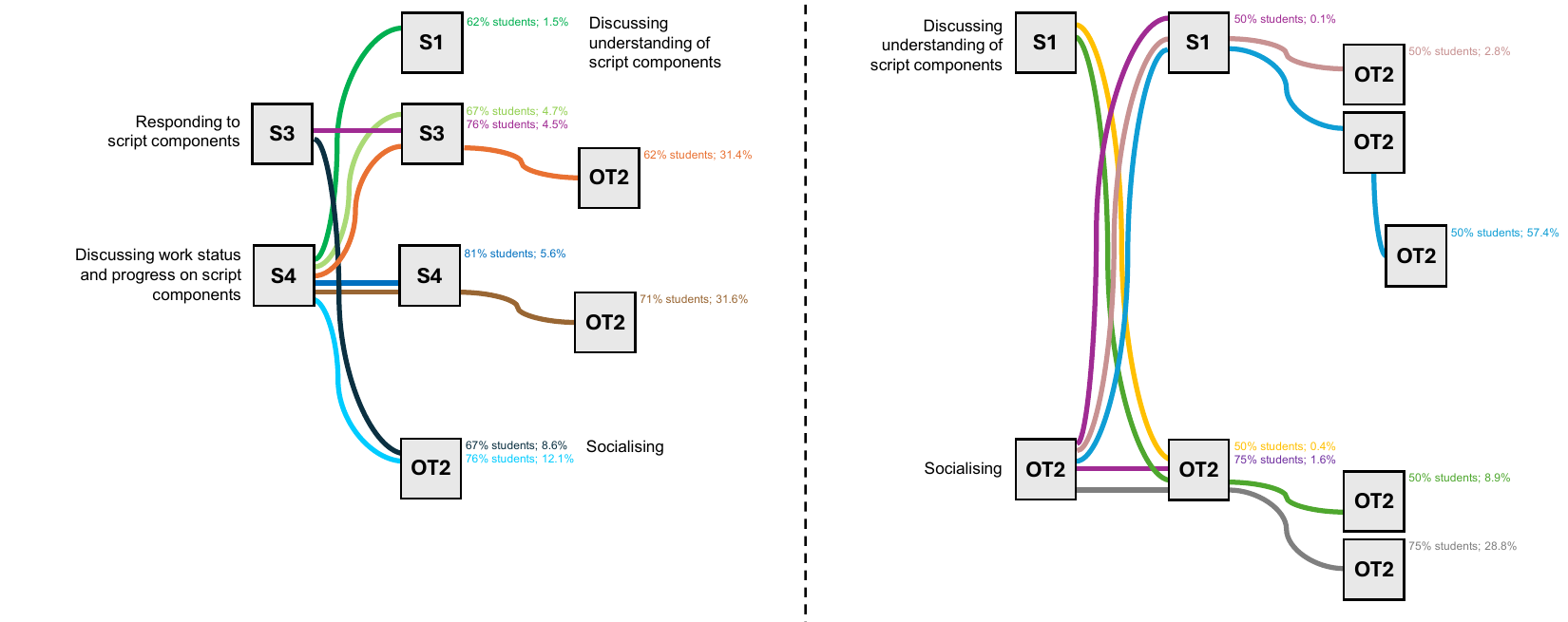}
\caption{Frequently occurring sequence of CPS indicators in maximal scaffolding among 21 students (left) compared to that in minimal scaffolding among 8 students (right) in the \textit{solution checking, problem extension and reflection} phase.}
\label{fig:spm_segment4}
\end{figure}

The visualisations of the flow diagrams for the subsequences produced using SPM in each CPS phase are shown in Figures \ref{fig:spm_segment1}, \ref {fig:spm_segment2}, \ref {fig:spm_segment3}, and \ref {fig:spm_segment4}. In Figure \ref{fig:spm_segment1}, we observe that there were no problem solving indicators observed in the subsequences for the maximal scaffold condition during the `\textit{problem identification}' phase. However, a problem-solving indicator (i.e., \textit{PS4}: asking questions to clarify understanding, ideas or contributions) was present in the subsequences for the minimal scaffold condition. Students progressed from this problem solving indicator to either \textit{PS4} again or \textit{S4} (i.e., discussing work status and progress on script components). It was also observed that in this phase, the number of frequently occurring sequences of CPS indicators in the minimal scaffold condition (i.e., 29 subsequences) was almost twice the number of subsequences in the maximal scaffold condition (i.e., 13 subsequences). Students in the minimal scaffold were engaging frequently in various sequences. In contrast, students in the maximal scaffold were more intentional and directly engaged in certain subsequences that involve the use of scripting indicators.

Although most of the subsequences in the maximal scaffolding involved scripting indicators (i.e., \textit{S1}, \textit{S3} and \textit{S4}), six subsequences involved students initially engaging in off-task indicators (\textit{OT1}: technical issues or logistical tasks related to the learning environment) but subsequently returning to scripting indicators. In the minimal scaffold condition, 9 subsequences involved students initially engaging in \textit{OT1} and one subsequence initially engaging in \textit{OT2} (i.e., socialising). Although all of these subsequences eventually ended with scripting indicators, five of these subsequences continued with an additional progression to OT1 before reaching scripting indicators. Maximal scaffold was observed to be sufficient in retaining students' focus and attention on the CPS task through the provided instructions. On the other hand, students with minimal scaffolding might lack sufficient support to progress away from off-task behaviours.

In the `\textit{ideation, planning and decision making}' phase (see Figure \ref {fig:spm_segment2}), we observe that subsequences in the maximal scaffold condition continued to not involve problem solving indicators but only scripting indicators (i.e., \textit{S1}, \textit{S3} and \textit{S4}). We also observe that these subsequences did not involve off-task indicators. Subsequences in the minimal scaffold condition, on the other hand, continued to include one problem solving indicator (\textit{PS}), with 6 out of 7 subsequences progressing to \textit{OT2} (i.e., socialising). The maximal scaffolding directed students' progression away from socialising behaviours. The decrease of instructional support in the minimal scaffold condition saw the progression of problem solving and scripting indicators to socialising behaviours. In this CPS phase, students must be able to gain perspective from each other to reach a consensus on the approach that they will take towards solving the given problem solving question. Engaging in such off-task behaviours can have negative impacts on students' learning from the CPS task \cite{Cocea2009}. 

In the `\textit{plan implementation and solution generation}’ phase (see Figure \ref {fig:spm_segment3}), we observe in the minimal scaffold condition that students progressed to the next problem solving indicator after engaging in scripting (e.g., \textit{S4} progressing to \textit{PS20}) or after demonstrating a problem solving indicator (i.e., \textit{PS4} progressing to \textit{PS3}, and \textit{PS20} progressing to \textit{PS26}). In particular, the identification of \textit{PS26} in the subsequence, where students were "discussing the progress and status of individual and team sub-tasks", was relevant in this CPS phase. However, in the maximal scaffold conditions, we observed that none of the subsequences involved problem solving indicators. Subsequences involving the progression from scripting indicators to \textit{OT2} were present in about 64.3\% of students' behavioural sequences. This observation evidenced the effects of overscripting, where students lost motivation after extensive script usage to engage in socialising activities unrelated to the task \cite{dillenbourg_2002Overscripting}. In this CPS phase, it would have been beneficial for students to receive the minimal scaffolding.

Finally, in the `\textit{solution checking, problem extension and reflection}' phase (see Figure \ref {fig:spm_segment4}), we observed that at least 50\% of the subsequences led to \textit{OT2} in both scaffold conditions. Despite students attempting to use and regulate the scripting that was provided, it did not lead to the demonstration of problem solving behaviours. Students may be having difficulty comprehending the instructional support provided in the scaffolds, or be unfamiliar with what needs to be done in this CPS phase. Solution checking and reflection involve metacognitive processes, and such skills are developed over extended periods of attention and time working in these processes \cite{Schoenfeld2016}. 

\section{Discussion}
\subsection{Implications on scaffold design and teaching}
Through the provision of maximal scaffolding, students demonstrated significantly higher quantities of dialogic engagement and diversity of dialogic engagement across the CPS phases (refer to Figure \ref{fig:boxplot}). This has implications for teaching practice, where students should be given detailed guidance on scripting activities to enable them to engage in all aspects of the CPS phases. Yet, it is apparent that the current designs in the maximal and minimal scaffolds are unable to address the gap in students' ability to engage metacognitively in the A4 CPS phase (i.e., \textit{solution checking, problem extension and reflection}). Intentional training of students' metacognition \cite{teong_2003effect} is needed to equip students with the competencies to apply the scripting.

It is also apparent from the HINA and SPM analysis that students do problem solving differently from that which was structured into the CPS task. These analyses shed light on the structure of the CPS process that the students take by identifying significant indicators and emergent subsequences (refer to Figure \ref{fig:hina_behavioural} and \ref{fig:spm_segment1}). Learning analytics with such information allows teachers to reflect on two critical questions. Firstly, why are particular problem solving behaviours not demonstrated or repeatedly demonstrated by students across the CPS phases? Secondly, how can the learning design be refined to encourage students to engage in problem solving behaviours across all CPS phases? This provides an opportunity for teachers to evaluate the scaffolding that was provided in the learning design of the CPS task. Beyond the performance outcomes that are achieved through the use of scaffolding, the two analytical methods presented in this study enable teachers to gain a robust description of the behaviour and engagement that a scaffold is designed to achieve. Through this, further investigation could examine how teachers could gradually reduce scaffolding over time. 

Maximal and minimal scaffolds affect the behaviour and engagement of students at different CPS phases, where each phase highlights different benefits and disadvantages of these scaffolds (see Figures \ref{fig:spm_segment1} - \ref{fig:spm_segment4}). Although the use of adaptive scaffolding could potentially mitigate the concern of overscripting caused by maximal scaffolding, studies investigating its potential have yet to examine it across the different CPS phases. While this could be largely due to the student demographic and the type of problem solving question that was considered in such studies, a challenge to providing adaptive scaffolding in CPS phases could be the lack of empirical data to decide what support should be adaptively provided in a particular CPS phase. For this, SPM analysis would be highly useful to determine the frequently occurring patterns of CPS sequences that would be associated with desired outcomes such as learning gains or task performance.

Yet, it is not the case that the provision of more information, such as detailed sequential patterns of CPS, is always useful. While the visualisation of the CPS subsequences is intended to provide an overview of students' progression from one indicator to another, the presentation of this information as learning analytics could be seen as an overload of information to teachers \cite{vanleeuwen_2015Learning}. On the contrary, as HINA uses statistical approaches to aggregate measures and determine significant associations, critical observations are retained while reducing the complexities involved for meaningful interpretation (see Figures \ref{fig:boxplot} and \ref{fig:hina_behavioural}). For example, HINA can be specifically leveraged to provide learning analytics that identify students who could be struggling or unwilling to engage across the CPS phases in the different scaffold conditions, by identifying the outliers that fall outside of the statistical bounds. SPM analysis and HINA can be complementary approaches, but should be leveraged differently based on the insights that are obtained.

\subsection{Comparison between HINA and SPM analysis}
HINA and SPM analyses are two different approaches to studying CPS processes. Here, we elaborate on how these two methods compare with one another to highlight the value that each brings towards understanding students' engagement and behaviour in the CPS process.

In HINA, we observe that students in the maximal scaffold condition were engaging extensively in problem solving indicators (i.e., \textit{PS4}, \textit{PS5}, \textit{PS6}, \textit{PS15}, and \textit{PS20}) during the \textit{problem identification} phase (see Figure \ref{fig:hina_behavioural}). However, through the use of SPM, we observe that there are no emergent subsequences involving the use of problem solving indicators (see Figure \ref{fig:spm_segment1}). HINA allows the identification of indicators that are significantly associated with the various CPS phases. This allows the identification of significant problem solving indicators that would otherwise be unnoticed from SPM analysis. One should be cautious to conclude that students do not significantly demonstrate particular behavioural indicators just because it is not observed in the SPM analysis, since dependencies between indicators are considered in the subsequences. At the same time, one should be mindful that CPS is a dynamic process, and that HINA alone does not provide elaboration on the progression from these uncovered problem solving indicators. A nuanced approach of leveraging both HINA and SPM analysis could provide complementary insights into what indicators could be considered significant and how they would progress to other indicators in the different phases of the CPS process.

At the same time, the difference in observation of indicators enables us to infer that even when students have significantly demonstrated particular problem solving indicators, their progression from these problem solving indicators is varied and infrequent. This has implications for teaching strategies in CPS, where teachers should not just provide information on how students can demonstrate particular CPS indicators but also provide guidance on how students can consistently move between CPS indicators. This could then strengthen pathways in the CPS process for students to have sustained CPS behaviours. 

Particularly, HINA provides aggregated information on students' diversity across the different CPS phases and the normalised quantity of dialogic engagement. Using these measures produced using HINA, we can pinpoint specific students who may fall outside of the statistical bounds of the cohort (i.e., $Q1 - 1.5 \times IQR$ to $Q3 + 1.5 \times IQR$, such that \textit{Q1} is the first quartile, \textit{Q3} is the third quartile, and \textit{IQR} is the inter-quartile range) (see Figure \ref{fig:boxplot}). SPM analysis is unable to provide such insights, but can determine the degree to which particular students have demonstrated certain subsequences. HINA would be appropriate to achieve the goal of providing a baseline for making comparisons across students (between-student analysis). SPM analysis enables fine-grained identification of specified subsequences in behavioural sequences of students (within-student analysis).

\subsection{Limitations and Future Work}
The observations of maximal scaffolding and minimal scaffolding could be limited to the context of the given study. Furthermore, some of the observed patterns could be largely stimulated by artefacts in the task design and emerging due to the granularity of coded CPS indicators, rather than reflecting natural learning processes. Future research examining the impact of these observed differences on students’ learning outcomes would need to involve more controlled causal investigations beyond the use of HINA and SPM, since both approaches are limited to correlational analysis.

While various other methodologies can be taken to investigate the structural effects of scaffolding (e.g., ENA and HMM), the explorations in this study aim to shed light on the complementary and contrasting approaches of HINA and SPM analysis. Through this, we argue for the value of HINA by illustrating observations that are otherwise not evidenced from SPM analysis. Future studies could investigate whether the insights derived from HINA represent unique observations that would not otherwise be obtained through alternate approaches.

It remains to be explored how these CPS processes are connected to students' learning outcomes. It would be valuable to understand how students' behaviour and processes could impact individual and group performance from CPS. This would provide additional inferential analysis of behaviour-phase differences and SPM pattern contrasts.

Finally, the results from the current study may be limited in their generalisability. This study provides a relevant and specific account of contrasting two scaffolds using two complementary approaches (i.e., HINA and SPM analysis). Future studies could adopt these methodologies to provide similar depths towards uncovering processes during CPS. 

\section{Conclusion}
This study makes distinct methodological and theoretical contributions to learning analytics and collaborative learning research. Theoretically, by employing the node-level measures in HINA, we found that maximal scaffolding significantly influenced not only the quantity but, more importantly, the diversity of student participation across CPS phases. Comprehensive scaffolding enables students to grasp the distinct learning opportunities designed into each stage of collaborative problem-solving more effectively. However, teachers should be mindful that maximal scaffolds can present challenges to learners' motivation and sustained engagement across all phases of the CPS phases.

Methodologically, we demonstrated the unique analytical value that HINA and SPM analysis contribute towards investigating CPS processes. By leveraging HINA's capacity to identify statistically significant edges in the constructed networks, we were able to delineate the specific behavioural indicators that students demonstrated in each CPS phase. By taking into account the dependencies between behavioural indicators, SPM analysis provides temporal insights into frequently occurring subsequences that are observed in each CPS phase. As such, HINA can be leveraged when the scope of study seeks to provide a baseline aggregate comparison across subjects involved in a given study (e.g., comparing engagement diversity of subjects in different group conditions). On the other hand, studies that are interested in examining the progression of states and deriving emergent patterns that qualitatively describe particular processes can benefit from the use of SPM analysis (e.g., identifying subsequences of CPS processes among students with higher performance outcomes).

Based on the results observed, this study clarified the complementary analytical strengths of HINA and SPM analysis. While SPM excels at analysing temporal sequences and studying how it changes over time, HINA focuses on modelling and analysing relationships between different types of entities (e.g., students, behaviours, phases). We demonstrated that heterogeneous interaction networks offer a unique methodological value by revealing the multi-entity structure of CPS processes, providing insights that are distinct from yet complementary to sequence-oriented approaches.

\appendix
\section*{Appendix.}
Diversity and quantity value of students, and subsequences from SPM are provided here: \url{https://osf.io/kdcsa/?view\_only=34f45a87da4f475fadb9e9b02ecea40f}

\bibliographystyle{splncs04}

\end{document}